\documentclass[10pt,pra,twocolumn,superscriptaddress,showpacs,floatfix]{revtex4}

\usepackage{graphicx}
\usepackage{amssymb}
\usepackage{times}
\usepackage{amsmath}
\usepackage{amsthm}
\usepackage{epstopdf}

\usepackage{hyperref}

\input epsf.tex
\usepackage{graphicx}
\usepackage{amsthm}

\begin{document}

\title{Continuous-variable measurement-device-independent quantum key distribution with \\ photon subtraction}

\author{Hong-Xin Ma}\affiliation
 {Zhengzhou Information Science and Technology Institute, Zhengzhou 450004, China
}\affiliation{Synergetic Innovation Center of Quantum Information and Quantum Physics, University of Science and Technology of China, Hefei, Anhui 230026, China}

\author{Peng Huang}\thanks{Corresponding author: huang.peng@sjtu.edu.cn}
\affiliation
 {State Key Laboratory of Advanced Optical Communication Systems and Networks, Shanghai Key Laboratory on Navigation and Location-based Service, and Center of Quantum Information Sensing and Processing, Shanghai Jiao Tong University, Shanghai 200240, China
}

\author{Dong-Yun Bai}\affiliation
 {State Key Laboratory of Advanced Optical Communication Systems and Networks, Shanghai Key Laboratory on Navigation and Location-based Service, and Center of Quantum Information Sensing and Processing, Shanghai Jiao Tong University, Shanghai 200240, China
}

\author{Shi-Yu Wang}\affiliation
 {State Key Laboratory of Advanced Optical Communication Systems and Networks, Shanghai Key Laboratory on Navigation and Location-based Service, and Center of Quantum Information Sensing and Processing, Shanghai Jiao Tong University, Shanghai 200240, China
}

\author{Wan-Su Bao}\affiliation
 {Zhengzhou Information Science and Technology Institute, Zhengzhou 450004, China
}\affiliation{Synergetic Innovation Center of Quantum Information and Quantum Physics, University of Science and Technology of China, Hefei, Anhui 230026, China}

\author{Gui-Hua Zeng}\affiliation
 {State Key Laboratory of Advanced Optical Communication Systems and Networks, Shanghai Key Laboratory on Navigation and Location-based Service, and Center of Quantum Information Sensing and Processing, Shanghai Jiao Tong University, Shanghai 200240, China
}

\begin{abstract}
It has been found that non-Gaussian operations can be applied to increase and distill entanglement between Gaussian entangled states. We show the successful use of the non-Gaussian operation, in particular, photon subtraction operation, on the continuous-variable measurement-device-independent quantum key distribution (CV-MDI-QKD) protocol.  The proposed method can be implemented based on existing technologies.
Security analysis shows that the photon subtraction operation can remarkably increase the maximal transmission distance of the CV-MDI-QKD protocol, which precisely make up for the shortcoming of the original CV-MDI-QKD protocol, and 1-photon subtraction operation has the best performance. Moreover, the proposed protocol provides a feasible method for the experimental implementation of the CV-MDI-QKD protocol.
\end{abstract}


\pacs{03.67.Hk, 03.67.-a, 03.67.Dd}
\maketitle

\section{Introduction}\label{Intr}
Quantum key distribution (QKD) [1] allows two distant authenticated users, Alice and Bob, to establish secure key in the presence of eavesdroppers. Quantum physics guarantees the unconditionally secure communication through unsecure channels. The existing QKD protocols have been mainly divided into the two following categories: discrete-variable (DV)QKD protocols [2-5] and continuous-variable (CV) QKD protocols [6-11]. The theoretical security of some representative protocols in these two categories of protocols has been fully analyzed [12-20].


Compared with DVQKD protocols (such as BB84 protocol), CVQKD protocols have the following advantages: First of all, the preparation of the light sources are relatively simple. Secondly, the detectors have low cost and high detection efficiency. Thirdly, they can be effectively compatible with existing optical communication systems. Last but not least, CVQKD protocols allow one to approach the ultimate limit of repeater-less communication, known as the PLOB bound [21].
For these reasons, CVQKD has received extensive attention and in-depth study in the research area of quantum communication.
The CVQKD protocol with Gaussian-modulated coherent states has been proved to be secure against collective attacks, both in asymptotic case [14] and finite-size regime [15], and its composable security has been fully proven [16-17].
Moreover, this protocol has been demonstrated both in laboratory [10,22] and field tests [23].
Furthermore, the longest secure transmission distance of CVQKD has been extended to 150 km [24], which meets the demand of metropolitan network and inter-city network.


Theoretically, QKD can be proved unconditional security However, in the process of practical experiment implementation, some assumptions in the security demonstration model can not be fully satisfied, which leads to some potential security vulnerabilities inevitably existing in the practical system [25-29]. A natural attempt to narrow the gap between theory and practice is to identify all security vulnerabilities hidden in the practical system, and provide appropriate remedial measures. Based on this idea, we should fully characterize each device as possible, and try to solve all the ``side channel" [30]. However, it is difficult to fully implement this operation. Inspired by the idea of entanglement swapping, two groups [31-32] proposed measurement-device-independent (MDI) QKD independently, where Alice and Bob are both senders and an untrusted third party is introduced to perform Bell-State Measurement (BSM) and communicate the corresponding result.
It needs to be emphasized that Ref. [31] solves the problem of side-channel attack in full generality, while Ref. [32] is limited to qubit systems.
MDI-QKD not only can remove all known or unknown side-channel attacks on detectors, but also has favorable performance and can be easily implemented in long distance experiments. Soon after MDI-QKD was put forward, it was well analyzed in theory [33-39] and demonstrated in experiments [40-42].

There are also two main approaches to implement MDI-QKD: discrete-variable (DV) MDI-QKD [32,43] and continuous-variable (CV) MDI-QKD [44-47]. Unfortunately, compared with that of DV-MDI-QKD, the maximal transmission distance of CV-MDI-QKD is unsatisfactory. Moreover, it is rather difficult to implement CV-MDI-QKD in experiment as the quantum efficiency of the existing detectors can not meet the experimental requirements at present [45].
Facing the shortcoming of the transmission distance of CV-MDI-QKD, many investigations are being dedicated to improving its performance. Interestingly, it has been demonstrated theoretically and
experimentally that the non-Gaussian operations, for instance, the photon subtraction and photon addition operations, can be used to increase and distill the entanglement in Gaussian entangled states [48-51], and thus to significantly enhance the transmission distance of the CVQKD protocols [52-54]. Attractively, the photon subtraction operation can be practically implemented with existing technologies. Inspired by the aforementioned advantages, which are analyzed in theory and subsequently demonstrated with simulations and experiments, in this paper, we propose a method to improve the performance of the CV-MDI-QKD protocol by using non-Gaussian operation, i.e.,the photon subtraction, which can be implemented under current technology. By performing a suitable photon subtraction operation with the entangled source located on Alice's side, the new protocol can lengthen the security transmission distance in contrast with the original protocols.  Furthermore, this idea provides a feasible scheme for the demonstration of the CV-MDI-QKD protocol in the experimental environment.

The paper is structured as follows. In Sec.~\ref{PS}, we first introduce the original CV-MDI-QKD protocol, then introduce the model of the CV-MDI-QKD protocol with photon subtraction. In Sec.~\ref{Cal},we derive the secret key of the CV-MDI-QKD protocol with photon subtraction. In Sec.~\ref{PerD},we give the numerical simulation and performance analysis. Conclusions are drawn in Sec.~\ref{Con}.
\begin{figure*}[htbp]
\resizebox{15cm}{!}{
\includegraphics{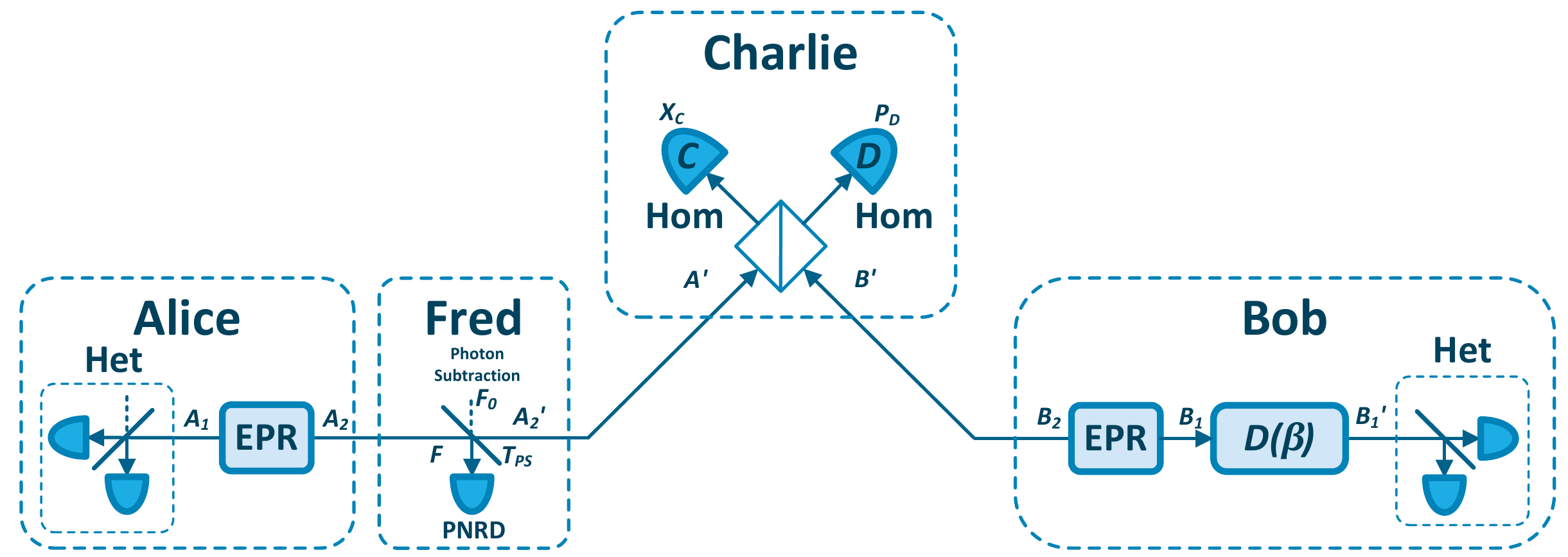}}
\caption{(Color online). EB scheme of the CV-MDI-QKD protocol with non-Gaussian operation, i.e.,the photon subtraction. EPR: two-mode squeezed state. Het: heterodyne detection. Hom: homodyne detection. PNRD: photon-number-resolving detector. \(D(\beta)\): displacement operation.}\label{fig:1}
\end{figure*}
\section{CV-MDI-QKD protocol with photon subtraction}\label{PS}  
In this section, we first review the original CV-MDI-QKD protocol, especially the EB scheme. Then, we present the model of CV-MDI-QKD protocol with photon subtraction (based on EB scheme).
\subsection{CV-MDI-QKD Protocol}

The EB scheme of the CV-MDI-QKD protocol is illustrated in part of Fig.~\ref{fig:1}. Alice generates one two-mode squeezed state with variance \(V_A\), where mode \(A_1\) is retained by Alice and mode \(A_2\) is sent to an untrusted third part Charlie through the quantum channel with length \(L_{AC}\). Bob generates another two-mode squeezed state with variance \(V_B\), where mode \(B_1\) is retained by Bob and mode \(B_2\) is sent to an untrusted third part Charlie through the quantum channel with length \(L_{BC}\). Charlie interferes two modes \(A'\) and \(B'\) at a beam splitter (BS) with two output modes \(C\) and \(D\). Then both the \(x\) quadrature of mode \(C\) and \(p\) quadrature of mode \(D\) are measured by Charlie through homodyne detection, and he announces the measurement results \(\left\{X_C,P_D\right\}\) through public channel.

After receiving \(\left\{X_C,P_D\right\}\), Bob modifies mode \(B_1\) to \(B'_1\) by displacement operation \(D(\beta)\), where \(\beta  = g\left( {{X_C} + i{P_D}} \right)\), and \(g\) represents the gain of the displacement operation. The density matrix of mode \(B'_1\) is
\begin{equation}
{\rho_{{{B'}_1}}} = D\left( \beta\right){\rho_{{B_1}}}{D}^\dag\left(\beta\right),
\end{equation}
where \(\rho_{B_1}\) is the density matrix of mod \(B_1\). Through these operations, mode \(A_1\) and \(B'_1\) become entangled [55]. By employing heterdyne detection, Bob get the quadratures of mode \(B'_1\): \(\left\{X_B,P_B\right\}\), and Alice get the quadratures of mode \(A_1\): \(\left\{X_A,P_A\right\}\). Obviously, \(\left\{X_B,P_B\right\}\) and \(\left\{X_A,P_A\right\}\) are correlated. Then  Alice and Bob implement information reconciliation and privacy amplification to obtaining a string of secret key. In addition, the total transmission distance \(L_{AB}\) is equal to \(L_{AC}+L_{BC}\).

\subsection{Photon subtraction in CV-MDI-QKD}
There are three ways to implement the CV-MDI-QKD protocol with photon subtraction: photon subtraction only located on Alice's side, photon subtraction only located on Bob's side,  and photon subtraction located on both sides.
In CV-MDI-QKD protocol, we suppose that Alice is the sender and Bob is the receiver, and Bob performs reverse reconciliation. Thus Alice's source is used to carry the information associated with secret key, and Bob's source, whose information is nearly removed, is used to assist in generating secret key. So when suitable photon subtraction is operated on Alice's side, the performance of the protocol will be improved. Reversely, when photon subtraction is operated on Bob's side, the performance has no improvement. Furthermore, the photon subtraction operation has a certain probability of failure, which will decrease the overall performance of the protocol with the photon subtraction operated on Bob's side. Taken together, the optimal scheme is the one that photon subtraction only located on Alice's side. Thus we mainly focus on the optical scheme where a photon subtraction operation, which is non-Gaussian operation, is located on Alice's side.

The EPR state \({\left|\psi\right\rangle_{{A_1}{A_2}}}\) is generated by squeezing two vacuum states with two-mode squeezed operator \(S(r)\), producing entangled mode \(A_1\) and \(A_2\) , where \(S(r) = exp[r({a_{A_1}}{a_{A_2}}-{a}_{A_1}^\dag {a}_{A_2}^\dag)/2]\), \(r\) is the squeezing parameter. Thus, the EPR state \({\left|\psi\right\rangle_{{A_1}{A_2}}}\) can be given by
\begin{equation}
{\left| \psi  \right\rangle _{{A_1}{A_2}}} = S(r)\left| {0,0} \right\rangle  = \sqrt {1 - {\xi^2}} \sum\limits_{n = 0}^\infty \xi^n {\left| {n,n} \right\rangle },
\end{equation}
where \(\xi  = \sqrt {\frac{{V_A - 1}}{{V_A + 1}}} \), \(V_A\) is the variance of the EPR state \({\left|\psi\right\rangle_{{A_1}{A_2}}}\), \({\left| {n,n} \right\rangle }  = {\left| n \right\rangle _{{A_1}}} \otimes {\left| n \right\rangle _{{A_2}}}\), and \(\left| {n} \right\rangle\) represents the Fock states.

In our protocol, we suppose that the photon subtraction operation is controlled by Fred, an untrusted third part which is near Alice. This kind of setting can lower requirements for the device perfection of the photon subtraction operation. The photon subtraction operation is shown in Fred's part of Fig.~\ref{fig:1}: Fred uses a BS with transmittance \(T_{PS}\) to split mode \(A_2\) and the vacuum state \(F_0\) into modes \(A'_2\) and \(F\). As \({\left| \psi  \right\rangle _{{A_1}{A_2}}}\) is a EPR state, mode \(A_1\) and \(A'_2\) are entangled, the tripartite state can be given by
\begin{equation}
{\rho _{{A_1}F{{A'}_2}}} = {U_{BS}}[{\left| \psi  \right\rangle _{{A_1}{A_2}}}{\left\langle \psi  \right|_{{A_1}{A_2}}} \otimes {\left| \psi  \right\rangle _{F_0}}{\left\langle \psi  \right|_{F_0}}]{U}^\dag _{BS},
\end{equation}
where \({\left| \psi  \right\rangle _{F_0}}={\left|0\right\rangle}\). Here we denote \({\rho _{{A_1}F{{A'}_2}}} =\left| \gamma  \right\rangle \left\langle \gamma  \right|\), then \({\left|\gamma \right\rangle}= {{{U_{BS}}{\left| \psi  \right\rangle _{{A_1}{A_2}}}\otimes {\left|0 \right\rangle}} }\). \({\left|\gamma \right\rangle}\) can be calculated as
\begin{equation}
\begin{array}{lll}
{\left|\gamma \right\rangle}=& {\sqrt {1 - {\xi ^2}} \sum\limits_{n = 0}^\infty  {\sum\limits_{m = 0}^n {{\xi ^n}\sqrt {C_n^m{{\left( {1 - {T_{PS}}} \right)}^m}{T_{PS}}^{n - m}} }}} \\ \\
                             &{\left| {n,m,n - m} \right\rangle},
\end{array}
\end{equation}
where \(n\) and \(m\) are natural number, \(n\ge m\), \(C_n^m\) is a combinatorial number. Mode \(F\) is measured by applying positive operator-valued measurement (POVM) \(\left\{ \rm M ^k_0, \rm M ^k_1\right\} \) [52] with employing the photo-number-resolving detector (PNRD), which can verify the photon number of quantum mode. The photon number of subtraction \(k\) depends on \({\rm M ^k_1} = \left| k \right\rangle \left\langle k \right|\). The photon subtraction operation is successful only when the POVM element \({\rm M ^k_1}\) clicks, and then the protocol can be proceeded.

The photon-subtracted state \(\rho _{{A_1}{{A'}_2}}^{PS}\) can be calculated as
\begin{equation}
\rho _{{A_1}{{A'}_2}}^{PS} = {P^{ - {\rm M^k_1}}}{\rm{t}}{{\rm{r}}_F}({\rm M^k_1}{\rho _{{A_1}F{{A'}_2}}}),
\end{equation}
where \({\rm{t}}{{\rm{r}}_\alpha }\left( \beta  \right)\) is the partial trace of mode \(\beta\) with tracing out its submode \(\alpha\). \({P^{{\rm M^k_1}}}\) is the normalization factor which denotes the success probability of subtracting \(k\) photons in mode \(A_2\). It is given by
\begin{equation}
\begin{array}{lll}
{P^{{\rm M^k_1}}}&={\rm {tr}}_{{{A_1}F{{A'}_2}}}({\rm M^k_1}{\rho _{{A_1}F{{A'}_2}}}) \\ \\
              &=(1 - {\xi ^2})\sum\limits_{n = k}^\infty  {C_n^k{\xi ^{2n}}{{(1 - {T_{PS}})}^k}{T_{PS}}^{n - k}} \\ \\
              &=(1 - {\xi ^2}){\left( {\frac{{1 - {T_{PS}}}}{{{T_{PS}}}}} \right)^k}\sum\limits_{n = k}^\infty  {C_n^k{{\left( {{\xi ^2}{T_{PS}}} \right)}^n}} \\ \\
              &=(1 - {\xi ^2}){\xi ^{2k}}\frac{{{{\left( {1 - {T_{PS}}} \right)}^k}}}{{{{\left( {1 - {\xi ^2}{T_{PS}}} \right)}^{k{\rm{ + }}1}}}},
\end{array}
\end{equation}
where \(n\ge k\). According to Eq.(5)(6), the covariance matrix of \(\rho _{{A_1}{{A'}_2}}^{PS}\) can be calculated by [57]
\begin{equation}
{\gamma^{PS}_{{A_1}{A'_2}}} = \left( {\begin{array}{*{20}{c}}
{X{{\rm I}_2}}&{Z{\sigma _z}}\\ \\
{Z{\sigma _z}}&{Y{{\rm I}_2}}
\end{array}} \right),
\end{equation}
where \({\rm I}_2\) is \(2 \times 2\) identity matrix, \({\sigma _z}=diag(1,-1)\), and
\begin{equation}
\begin{array}{l}
X = \frac{{2\left( {1 + k} \right)}}{{1 - {\xi ^2}{T_{PS}}}} - 1,\\ \\
Y = \frac{{2\left( {1 + k{\xi ^2}{T_{PS}}} \right)}}{{1 - {\xi ^2}{T_{PS}}}} - 1,\\ \\
Z = \frac{{\sqrt {{T_{PS}}} \xi \left( {1 + k} \right)}}{{1 - {\xi ^2}{T_{PS}}}}.
\end{array}
\end{equation}

After the photon subtraction operation, the bipartite state \(\rho _{{A_1}{A'_2}}^{PS}\) in Eq. (5) is not Gaussian anymore. However, by introducing the non-Gaussian operation, the entanglement degree of this state is increased [52].

\section{Calculation of the secret key rate}\label{Cal}

In this section, we mainly focus on the secret key rate of our protocol under one-mode collective Gaussian attack, where Bob performs reverse reconciliation. Through a photon subtraction operation, a original Gaussian state will turn into a non-Gaussian state. We suppose the secure key rate of CV-MDI-QKD protocol with photon subtraction is \(K_{PS}\), and \(K\) is the secure key rate of the one without photon subtraction, whose source is Gaussian state. According to the optimality of Gaussian attack [58], \(K_{PS}\) is no less than \(K\). Then the lower bound of \(K_{PS}\) can be estimated by using similar covariance matrix of Gaussian state.

\begin{figure}[!h]\center
\resizebox{9cm}{!}{
\includegraphics{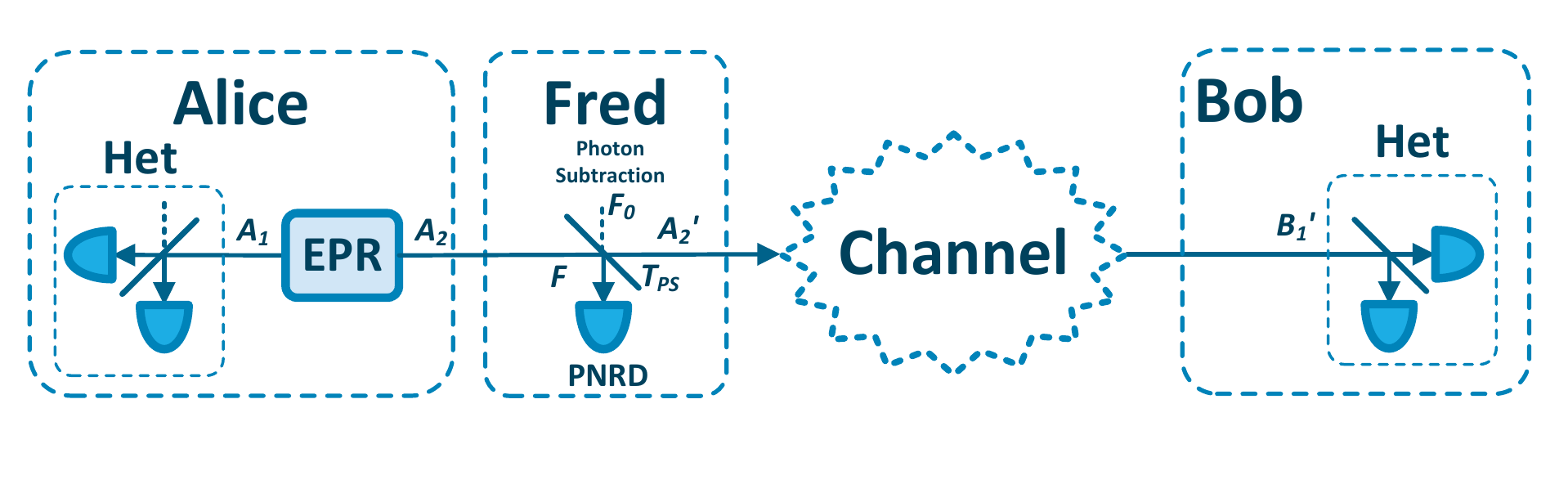}}
\caption{(Color online). Equivalent one-way protocol of the CV-MDI-QKD protocol with photon subtraction (EB scheme), where Eve is aware of Charlie and all Bob's operations except the heterdyne detection.}\label{fig:2}
\end{figure}

In Fig.~\ref{fig:1}, we assume that all Bob's operations except the heterdyne detection are untrusted, then the EB scheme of CV-MDI-QKD protocol is converted into a common one-way CVQKD protocol using heterdyne detection, which is shown in Fig.~\ref{fig:2}. Thus the CV-MDI-QKD protocol can be regarded as a specific case of the one-way CVQKD protocol which Eve obtains less information. Suppose the secure key rate of the equivalent one-way protocol is \(K_{one}\), then we can figure out that \(K_{PS} \ge K_{one}\). Through calculating the secret key rate of equivalent one-way protocol by using similar covariance matrix of Gaussian state, we can give the lower bound of \(K_{PS}\). For the convenience of analysis, we use the lower bound of \(K_{PS}\) to calculate the secure key rate of the CV-MDI-QKD protocol with photon subtraction.

The CV-MDI-QKD protocol has two quantum channels, which is different from the one-way protocol. There are two ways for Eve to eavesdropping: taking entangling cloner attacks on each quantum channel independently, which is called one-mode attack, or taking correlated two-mode coherent Gaussian attack where Eve injects quantum correlations in both quantum channels, which is called two-mode attack. When the two quantum channels are not associated, the two-mode attack degenerates into the one-mode attack. Therefore, the two-mode attack is a more general attack model. In Ref. [36] and [46], the CV-MDI-QKD protocol against two-mode coherent Gaussian attack is well analyzed, which is demonstrated to be the optimal attack strategy.

In the practical system, when the two quantum channels come from different directions, the correlation between the ambient noise of the two quantum channels should be very weak. Besides, there are some technical difficulties for Eve to inject quantum correlations in both quantum channels. Therefore, we restrict our analysis to two Markovian memoryless Gaussian quantum channels, which do not interact with each other. Then, the quantum channels of CV-MID-QKD protocol can be reduced to a one-mode channel [59]. Under this condition, the optimal attack strategy degenerates into the one-mode attack, and taking entangling cloner attacks on each quantum channel independently can be taken as the optimal one-mode collective Gaussian attack. The following secret key rate calculation, simulation and discussion are based on one-mode collective Gaussian attack. In addition, we should point out that Eve¡¯s attack described here is not the optimal one.

We assume the thermal excess noise and the transmittance of the quantum channel between Alice (Bob) and Charlie are \(\varepsilon_A^{th}\) (\(\varepsilon_B^{th}\)) and \(T_A\) (\(T_B\)). The transmittance can be given as
\begin{equation}
{T_A} = {10^{\frac{{ - l {L_{AC}}}}{{10}}}},{T_B} = {10^{\frac{{ - l {L_{BC}}}}{{10}}}},
\end{equation}
where \(l\) = 0.2 dB/km is both quantum channel losses. Here we set a normalized parameter \({\rm{T = }}\frac{{{T_A}{g^2}}}{2}\) is  which is associated with transmittance of quantum channel, \(g\) is the gain of the displacement operation in Bob. \(\chi _{line}=\frac{{1 - {\rm{T}}}}{\rm{T}} + \varepsilon^{th}\) is the total channel-added noise expressed in shot noise units. \(\varepsilon^{th}\) refers to the equivalent thermal excess noise of the equivalent one-way protocol, which is shown in Fig.~\ref{fig:2}. \(\varepsilon^{th}\) can be calculated by
\begin{equation}
\begin{array}{lll}
\varepsilon^{th}  =& \frac{{{T_B}}}{{{T_A}}}{{\left( {\sqrt {\frac{2}{{{T_B}{g^2}}}}\sqrt {{V_B}-1}  - \sqrt {{V_B} + 1} } \right)}^2}\\ \\
&+\frac{{{T_B}}}{{{T_A}}}{\left( {{\chi _B} - 1} \right)} +{\chi _A}+1,
\end{array}
\end{equation}
where \({\chi _A} = \frac{1}{{{T_A}}} - 1 + {\varepsilon _A^{th}}, {\chi _B} = \frac{1}{{{T_B}}} - 1 + {\varepsilon _B^{th}}\), \(V_B\) is the modulation variance of the EPR state in Bob. In order to minimize \(\varepsilon^{th}\), we adopt \({g^2} = \frac{{2({V_B-1}) }}{{{T_B}\left( {{V_B} + 1} \right)}}\), then
\begin{equation}
\begin{array}{lll}
\varepsilon^{th}  &= \frac{{{T_B}}}{{{T_A}}}\left( {{\chi _B} - 1} \right) + 1 + {\chi _A} \\ \\
 &= \frac{{{T_B}}}{{{T_A}}}\left( {{\varepsilon _B^{th}} - 2} \right) + {\varepsilon _A^{th}} + \frac{2}{{{T_A}}}.
\end{array}
\end{equation}

Since the homodyne detectors in Charlie are not the ideal apparatuses, the detection-added noise can be expressed in shot noise units as \({\chi _{hom}} = \left[ {\upsilon _{el}  + \left( {1 - \eta } \right)} \right]/\eta \), where \(\eta\) is the quantum efficiency of the homodyne detector, \(\upsilon _{el}\) is the electronic noise of the detector. The total noise referred to the channel input is expressed as \({\chi _{t}} = {\chi _{line}} + 2{\chi _{hom}}/{\rm{T}}\).

When Bob performs reverse reconciliation, the secure key rate of the CV-MDI-QKD protocol with photon subtraction under one-mode collective Gaussian attack is calculated as

\begin{equation}
{K_{PS}} = {P^{{\rm M^k_1}}}(\beta {I^{PS}_{AB}} - {\chi^{PS} _{BE}}),
\end{equation}
where \(I^{PS}_{AB}\) is the Shannon mutual information between Alice and Bob, \(\chi^{PS} _{BE}\) are the Holevo bounds [60] between Bob and Eve, which put an upper limit on the information available to Eve on Bob's key. \(\beta \) is the reconciliation efficiency for reverse reconciliation, \({P^{{\Pi ^k}}}\) is the success probability of \(k\)-photon subtraction operation. Detailed calculation of the secret key rate can be found in Appendix A.

\begin{figure}[!h]\center
\resizebox{8cm}{!}{
\includegraphics{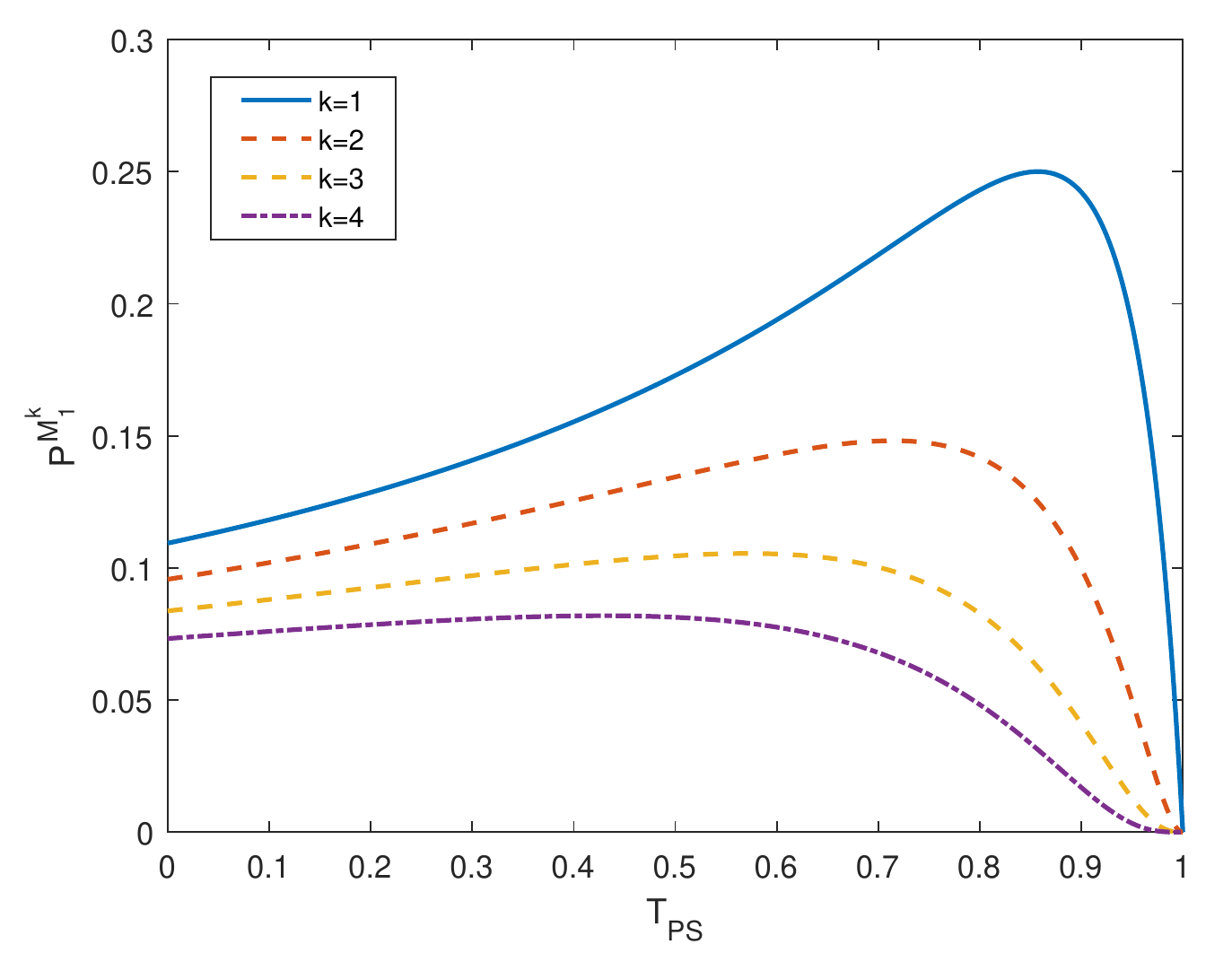}}
\caption{(Color online). Relation of the success probability of subtracting \(k\) photons and transmittances of BS in photon subtraction operation. The variance of EPR state is \(V_A=V_B=15\).}\label{fig:3}
\end{figure}

\begin{figure}[!h]\center
\resizebox{8cm}{!}{
\includegraphics{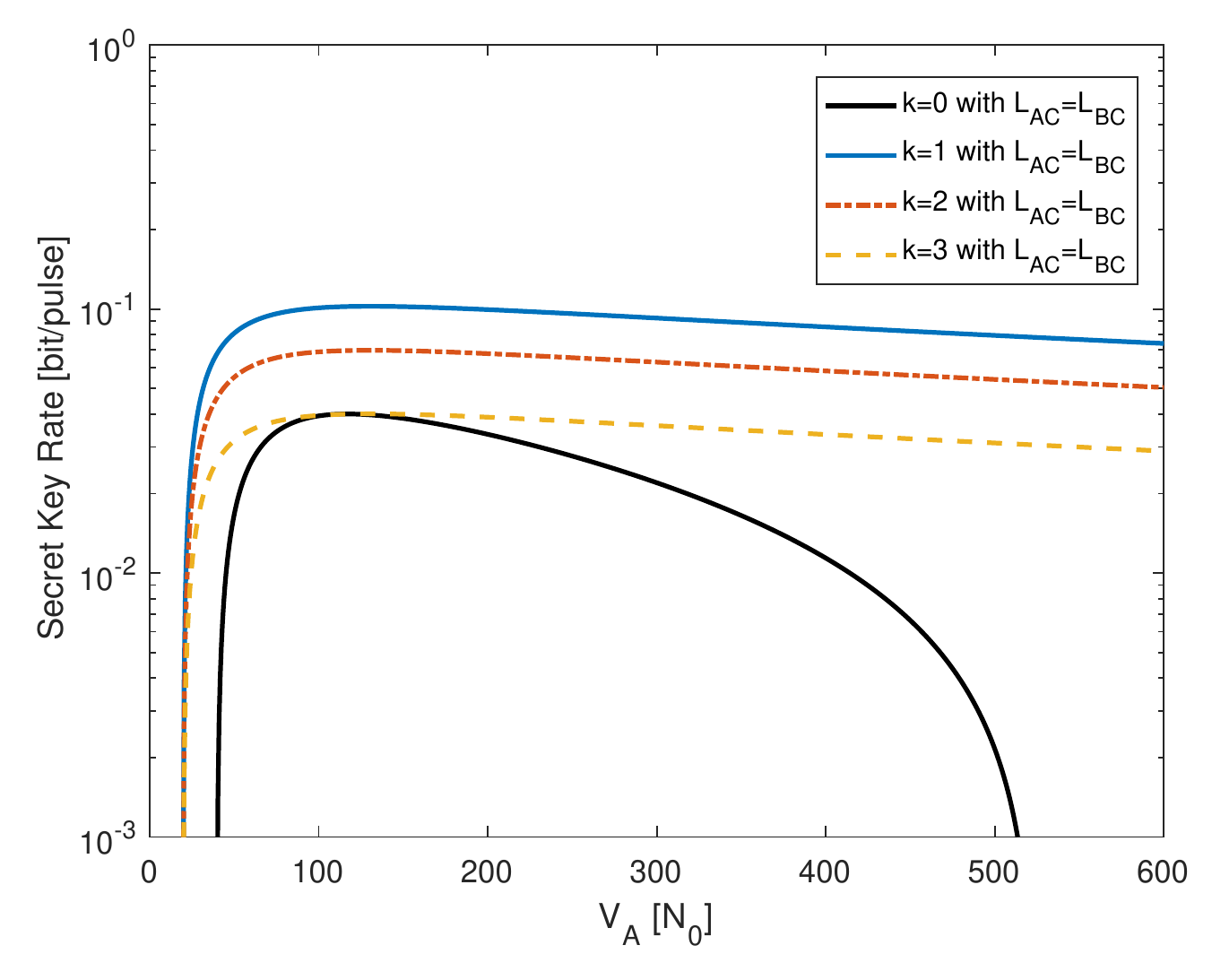}}
\caption{(Color online). The secret key rates as a function of \(V_{A}\) for every optimal \(T_{PS}\) in the symmetric case, where \(L_{AC}\) = \(L_{BC}\) and transmission distance \(L_{AB}\) = 6 km. \(N_0\) is the shot noise variance. Other parameters are fixed as follows: thermal excess noise \(\varepsilon _A^{th}=\varepsilon _B^{th}=0.01\), quantum efficiency of homodyne detector \(\eta=0.975\), electronic noise of homodyne detector \(\upsilon _{el}=0.01\) and reconciliation efficiency \(\beta=96\%\).}\label{fig:4}
\end{figure}

\section{Performance analysis and Discussion}\label{PerD}
In this section, we show the performance of CV-MDI-QKD protocol with photon subtraction compared with the original CV-MDI-QKD protocol.

First of all, we give the relation of the success probability of subtracting \(k\) photons and transmittances of BS in photon subtraction operation with different photon number of subtraction, which is based on Eq.6 and shown in Fig.~\ref{fig:3}. The success probability \(P^{{\rm M^k_1}}\) should be careful considered as it is of vital importance in calculating the secret key rate. Higher success probability will lead to higher secure key rate and further transmission distance.
As illustrated in Fig.~\ref{fig:3}, for different photon number of subtraction, there is always a specific \(T_{PS}\) leads to the optimal value of \(P^{{\rm M^k_1}}\). We observe that the optimal success probability is reduced with increasing photon number of subtraction. By subtracting more photons, more noise is added to the covariance matrix [61], which leads to the worse performance.
Note that for all the CV-MDI-QKD protocols with photon subtraction, 1-photon subtraction operation can obtain the highest success probability.

\begin{figure}[!h]\center
\resizebox{8cm}{!}{
\includegraphics{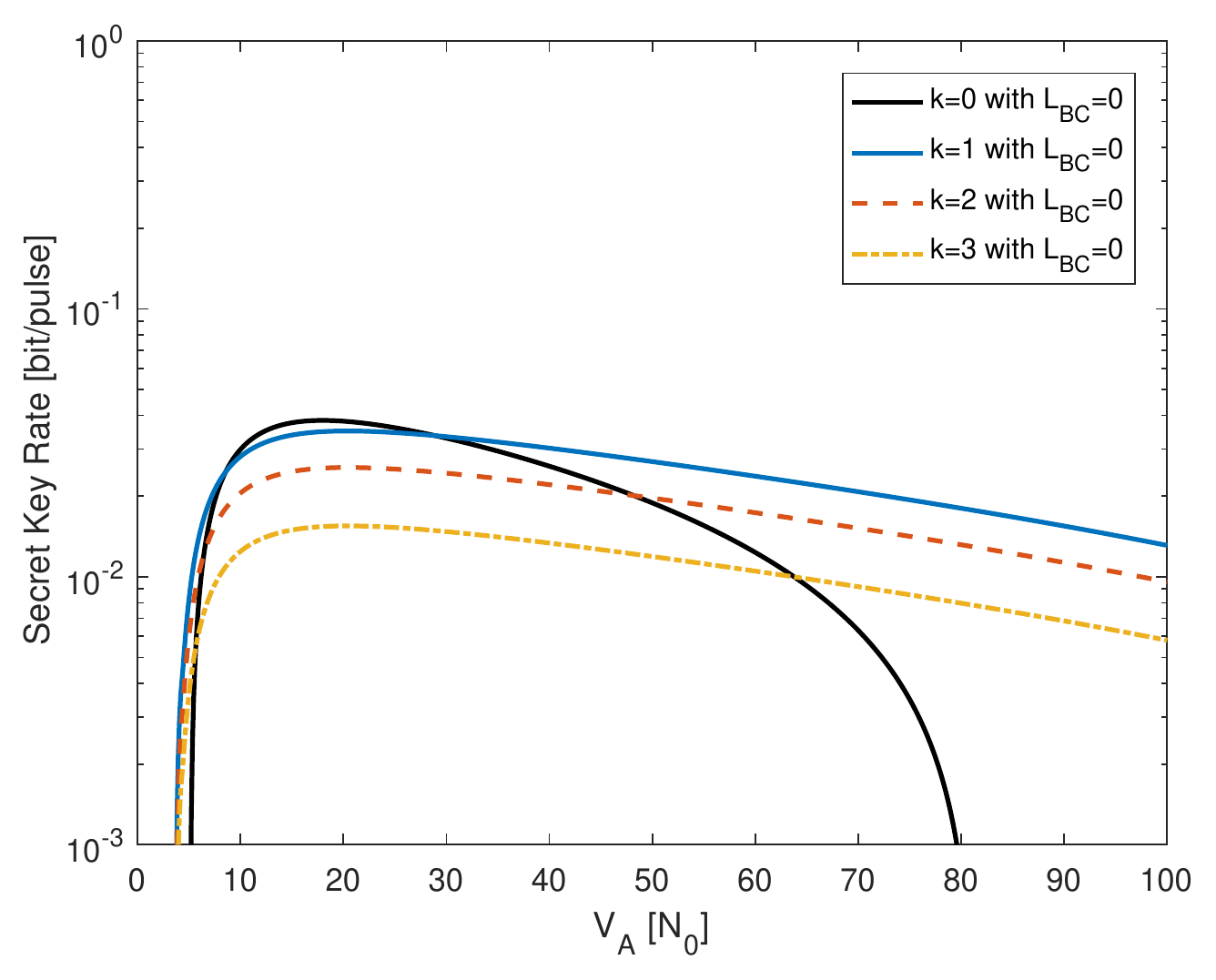}}
\caption{(Color online). The secret key rates as a function of \(V_{A}\) for every optimal \(T_{PS}\) in the extreme asymmetric case, where \(L_{BC}=0\) and transmission distance \(L_{AB}\) = \(L_{AC}\) = 30 km. Other parameters are fixed as same as Fig.~\ref{fig:4}.}\label{fig:5}
\end{figure}

The variance \(V_{A}\) and \(V_{B}\) are of vital importance in CV-MDI-QKD protocol. They determine the transmitting power of the quantum signal and finally affects the transmission distance and the secret key rate of the whole system. In this protocol, we set \(V_{A}=V_{B}\), and the following analysis only for \(V_{A}\).
In practice, the statistical variance of the practical prepared EPR state is in error with \(V_{A}\). The larger the optimal areas of \(V_{A}\), the higher the flexibility and stability of the system. Thus, we need to focus on the optimal value and optimal areas of \(V_{A}\). The plot of Fig.~\ref{fig:4} and Fig.~\ref{fig:5} show the secret key rates as a function of \(V_{A}\) in symmetric case (\(L_{AC}=L_{BC}\)) and extreme asymmetric case (\(L_{BC}=0\)), respectively. For the convenience in analysis, we set the transmission distance is 6 km in symmetric case and 30 km in extreme asymmetric case. In symmetric case, the secret key rate is at the peak value when the modulation variance \(V_{A}\) is about 100, namely the optimal value of \(V_{A}\) is about 100. In extreme asymmetric case, the optimal value of \(V_{A}\) is about 15.
Compared with the original CV-MDI-QKD protocol, the secret key rate of the protocols employing photon subtraction operation decrease much more slowly with the increase of \(V_{A}\), where the \(V_{A}\) is greater than the optimal value, and the protocol employing 1-photon subtraction operation has the slowest tempo of decrease. In this way, the proposed protocols have much larger optimal areas of \(V_{A}\) than the original protocol, which means that the proposed protocols can have a more flexible application , and the protocol employing 1-photon subtraction operation is most flexible. Moreover, as the SNR will be improved with the increase of variance, the proposed protocols can improve the performance by appropriately increase variance \(V_{A}\), whose feasible maximum value is far bigger than the one of the original protocol.
To summarise, the introduction of photon subtraction operation can obviously improve the flexibility and stability of the CV-MDI-QKD protocol, and the application of 1-photon subtraction operation works best.

\begin{figure}[!h]\center
\resizebox{8cm}{!}{
\includegraphics{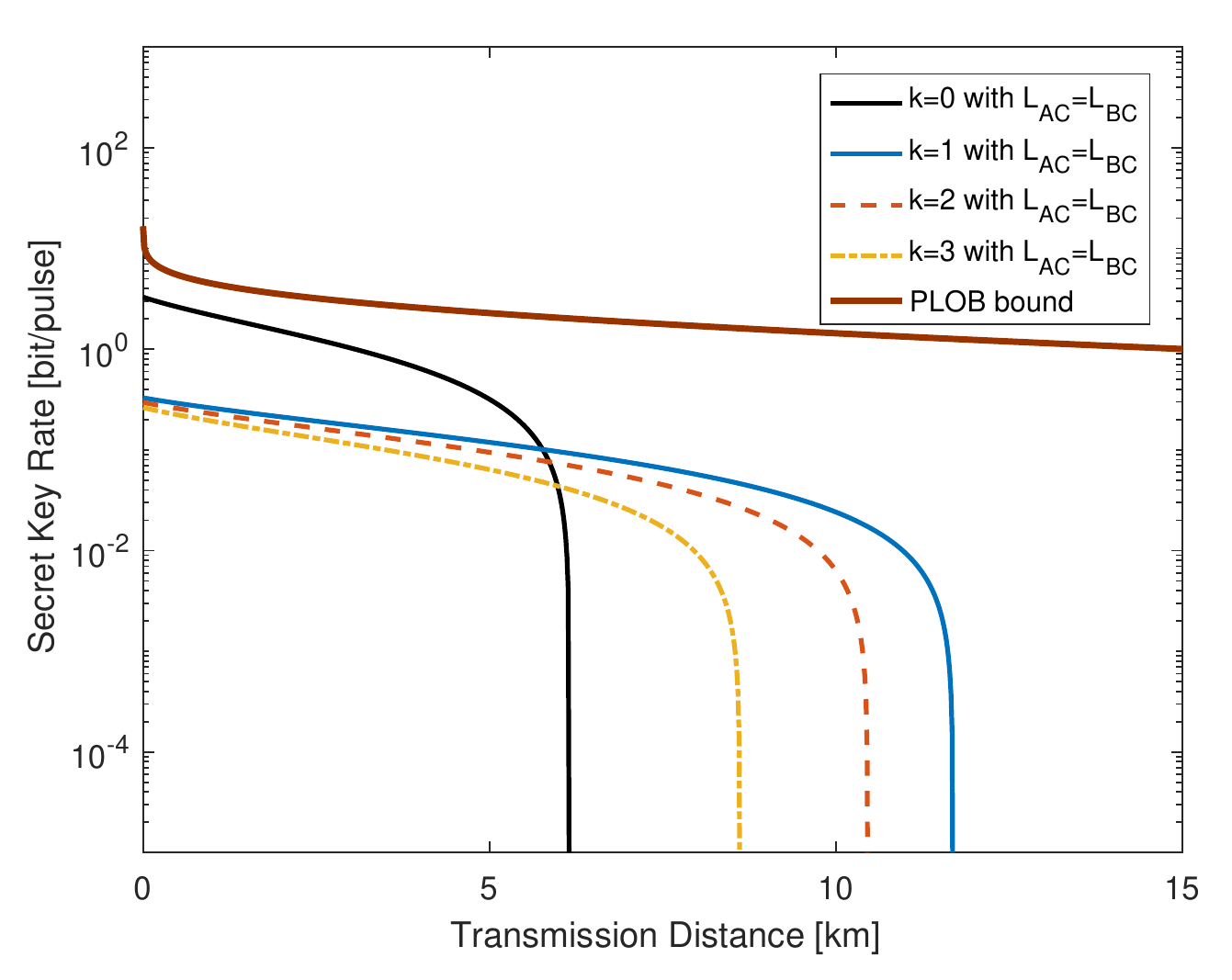}}
\caption{(Color online). The secret key rates as a function of the transmission distance from Alice to Bob for every optimal \(T_{PS}\) in the symmetric case, where \(L_{AC}=L_{BC}\). The variance of EPR state is \(V_A=V_B=100\). Other parameters are fixed as follows: thermal excess noise \(\varepsilon _A^{th}=\varepsilon _B^{th}=0.01\), quantum efficiency of homodyne detector \(\eta=0.975\), electronic noise of homodyne detector \(\upsilon _{el}=0.01\) and reconciliation efficiency \(\beta=96\%\).}\label{fig:6}
\end{figure}

The plot of Fig.~\ref{fig:6} shows the secret key rates as a function of the transmission distance from Alice to Bob, where \(T_{PS}\) takes the optimal values for each \(k\). The numerical simulation is given in the symmetric case where \(L_{AC}=L_{BC}\). The solid line labeled as \(k=0\) denotes the secret key rate of the original CV-MDI-QKD protocol, whose performance is worse than the CV-MDI-QKD protocols with photon subtraction in usual transmission distance. In other words, the photon subtraction operation can effectively enhance the secret key rate in usual case and extend the maximum transmission distance. Among all the CV-MDI-QKD protocols with photon subtraction,the protocol with 1-photon subtraction operation has the best performance, which can obtain longest maximum transmission distance.

\begin{figure}[!h]\center
\resizebox{8cm}{!}{
\includegraphics{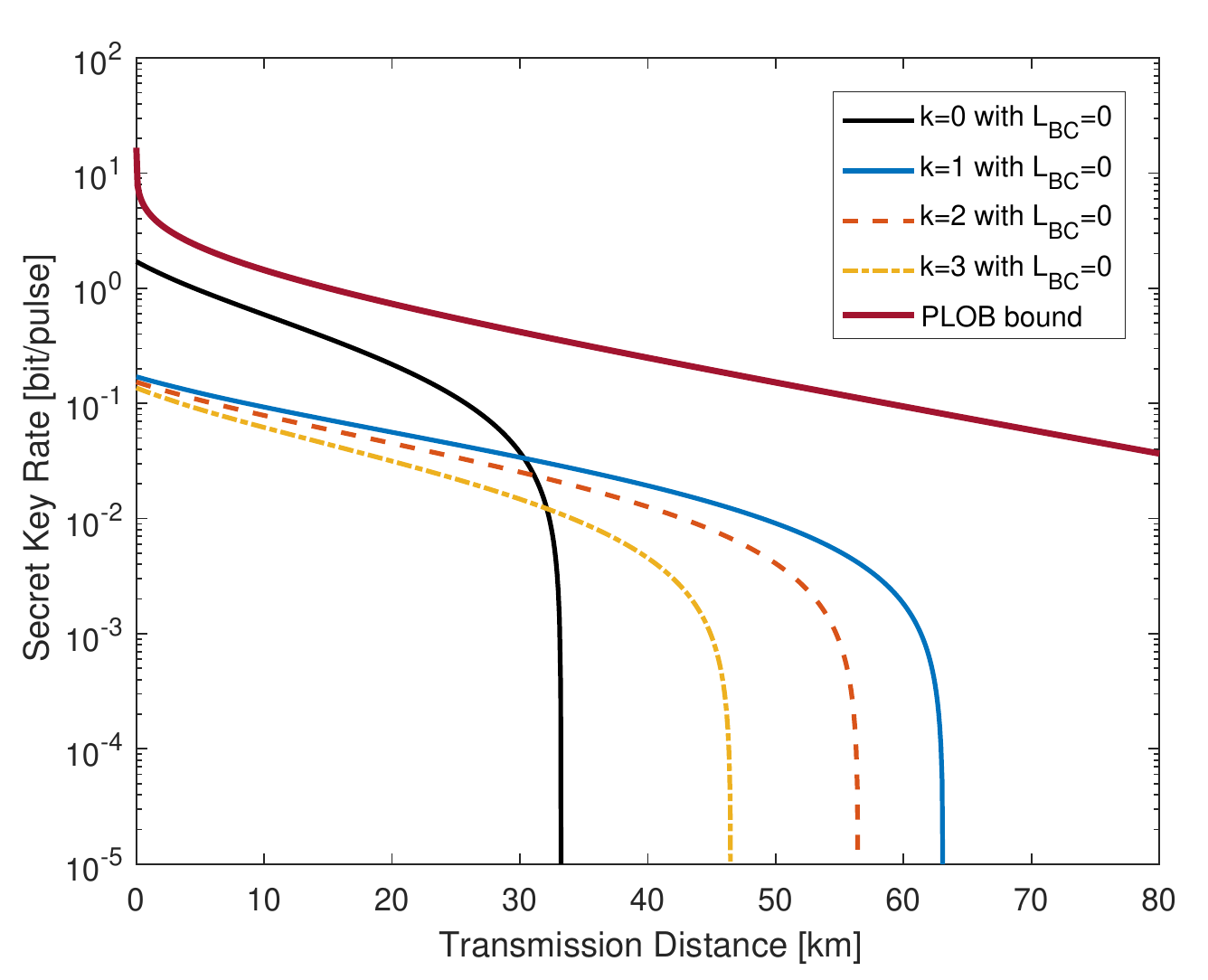}}
\caption{(Color online). The secret key rates as a function of the transmission distance from Alice to Bob for every optimal \(T_{PS}\) in the extreme asymmetric case, where \(L_{BC}=0\). The variance of EPR state is \(V_A=V_B=15\). Other parameters are fixed as same as Fig.~\ref{fig:6}.}\label{fig:7}.
\end{figure}

Ref. [46] has found that the optimal network configuration is the the extreme asymmetric case, where the untrusted third part Charlie acts as a proxy server near to one of the legitimate parties. Here we set Charlie is extreme close to Bob, where \(L_{BC}=0\). Then the effective transmission distance is equal to \(L_{AC}\). The secret key rates as a function of the transmission distance in this situation is plotted in Fig.~\ref{fig:7}. The maximum transmission distance of original CV-MDI-QKD protocol can be a relatively longer distance, up to 33.2 km is theory. The protocols with photon subtraction are still have longer maximum transmission distance than the original protocol, and the protocol with 1-photon subtraction operation has the best performance, of which the maximum transmission distance has reached 63 km in theory.

In addition, the PLOB bound has been plotted in both Fig.~\ref{fig:6} and Fig.~\ref{fig:7}, which shows the ultimate limit of repeater-less communication. By contrast, we can easily find that, when the transmission distance is larger than 5.7 km in the symmetric case (or 30.6 km in the extreme asymmetric case), the performance of the CV-MID-QKD protocol with 1-photon subtraction is closer to the PLOB bound than that of the original CV-MDI-QKD protocol. However, both of them cannot outperforms the PLOB bound at any transmission distance,and the gap between them and PLOB bound will become larger and larger as the transmission distance increases. Furthermore, the MDI node may act as active repeater able to beat the PLOB bound which refers to repeater-less point-to-point communications.

Yet there is a case worthy of our attention: in the very short transmission distance range, the secret key rate of the CV-MDI-QKD protocols with photon subtraction is worse than that of original CV-MDI-QKD protocol. This phenomenon is caused by the low successful probability of photon subtraction operation in low-channel-loss case.

By properly taking advantage of photon subtraction operation, the maximal transmission distance of CV-MDI-QKD protocol have risen by 91.8 per cent in the symmetric case and 89.7 per cent in the extreme asymmetric case. The effect of performance improvement is considerable, especially for the significant limitations of the CV-MDI-QKD protocol over the transmission distance.

\begin{figure}[!h]\center
\resizebox{9cm}{!}{
\includegraphics{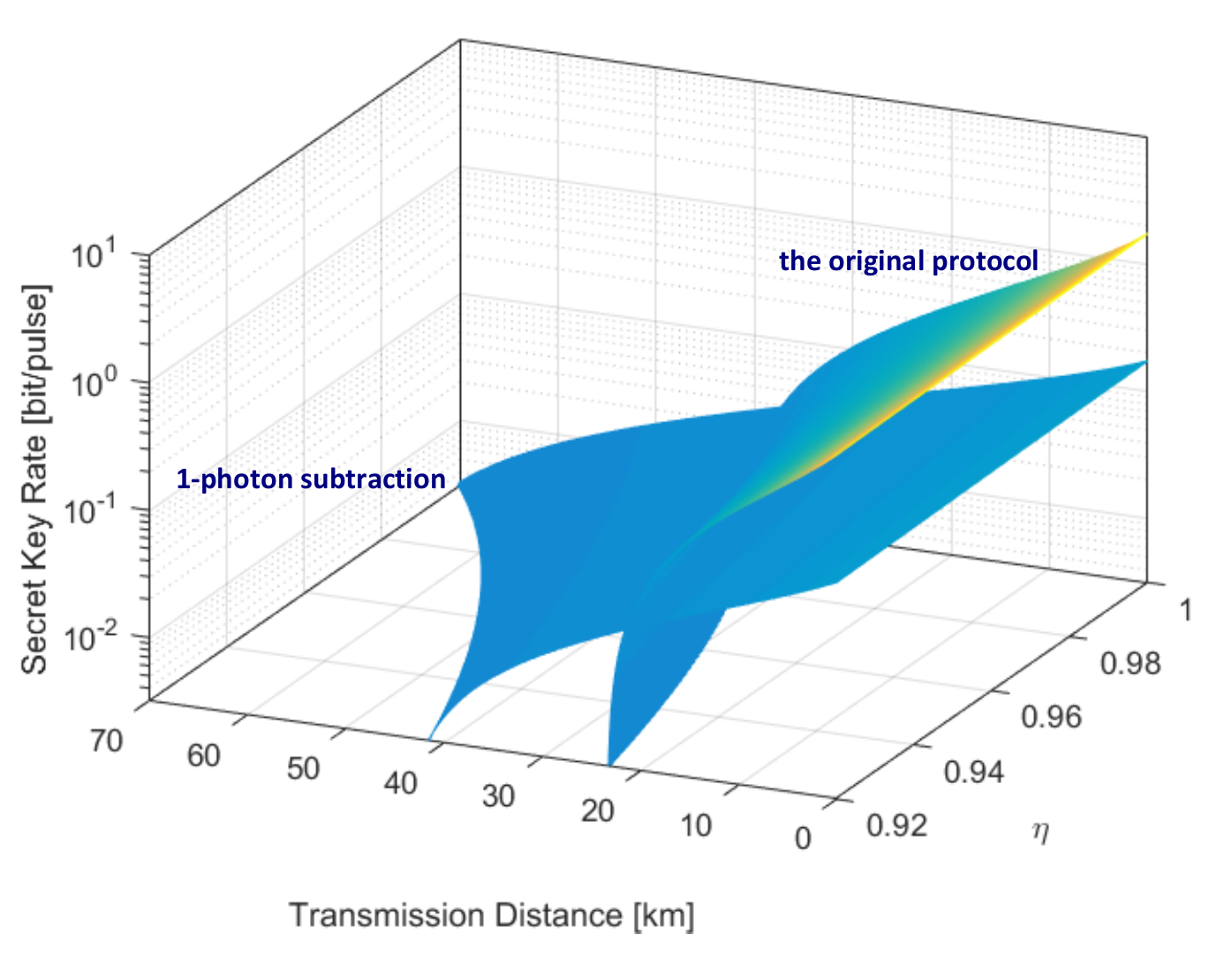}}
\caption{(Color online). Performance comparison of the CV-MID-QKD protocol with 1-photon subtraction and the original CV-MDI-QKD protocol for different quantum efficiency of homodyne detector in the extreme asymmetric case. Parameters are fixed as follows: the variance of EPR state is \(V_A=V_B=15\), thermal excess noise \(\varepsilon _A^{th}=\varepsilon _B^{th}=0.01\), electronic noise of homodyne detector \(\upsilon _{el}=0.01\) and reconciliation efficiency \(\beta=96\%\).}\label{fig:8}.
\end{figure}

Until now, there is no experimental feasible CV-MDI-QKD protocol has been proposed. The biggest difficulty is that the quantum efficiency of existing detectors can not meet the requirements of the protocol implementation [45]. Fig.~\ref{fig:8} depicts the performance comparison of the CV-MID-QKD protocol with 1-photon subtraction and the original CV-MDI-QKD protocol for different quantum efficiency of homodyne detector in the extreme asymmetric case. Under the framework of metropolitan area, where the transmission distance is larger than 30.6 km, the performance of the CV-MID-QKD protocol with 1-photon subtraction is always better than that of the original CV-MDI-QKD protocol when the quantum efficiency of homodyne detector is a definite value. In other words, the CV-MID-QKD protocol with 1-photon subtraction requires lower detector's quantum efficiency than the original CV-MDI-QKD protocol in the case of achieving the same performance. Adding 1-photon subtraction operation makes the CV-MDI-QKD protocol more tolerant of the quantum efficiency of homodyne detector in metropolitan area.

\begin{figure}[!h]\center
\resizebox{8cm}{!}{
\includegraphics{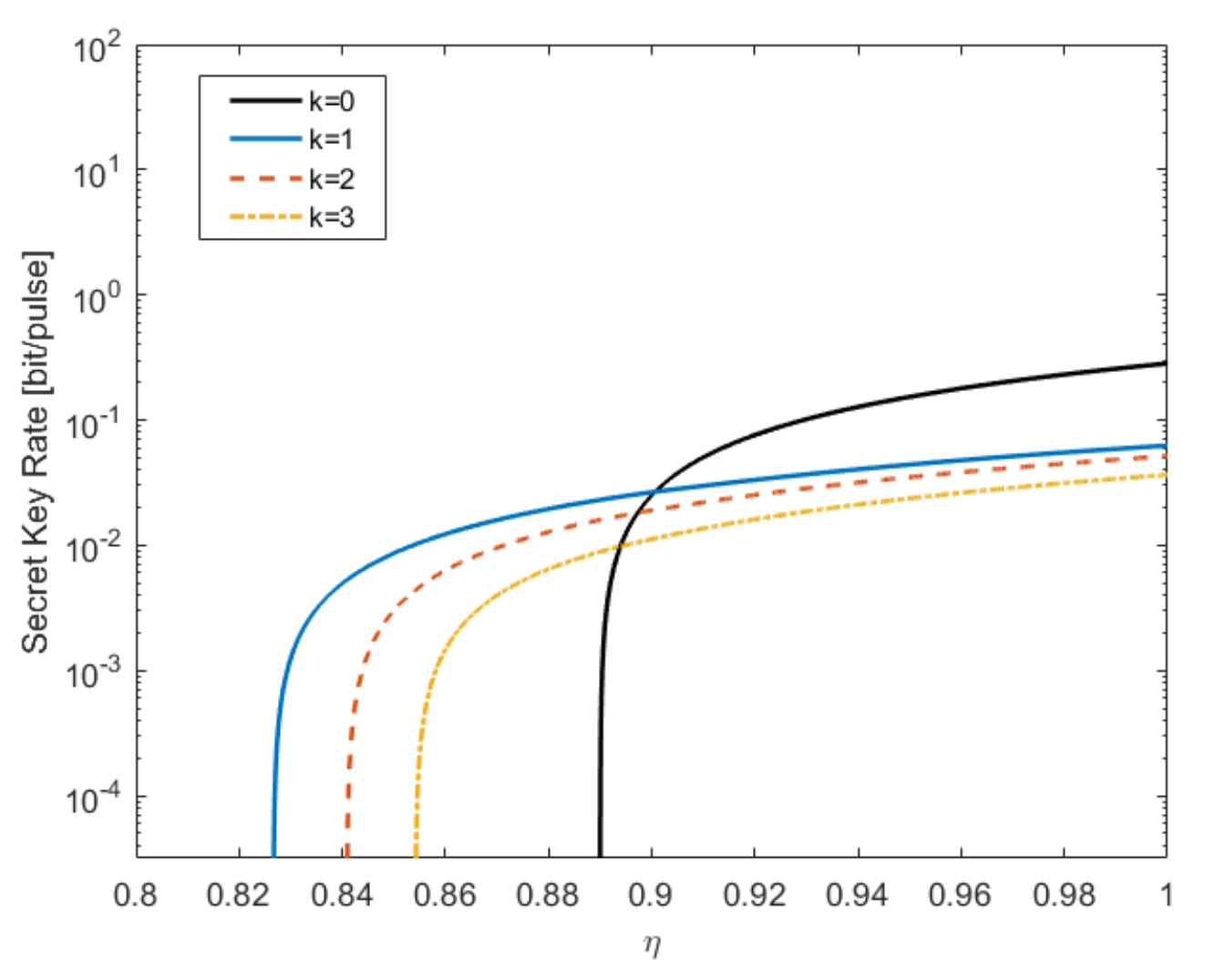}}
\caption{(Color online). The secret key rates as a function of the quantum efficiency of homodyne detector in the extreme asymmetric case. The transmission distance is 20 km. Other parameters are fixed as same as Fig.~\ref{fig:8}. }\label{fig:9}.
\end{figure}

Fig.~\ref{fig:9} shows the secret key rates as a function of the
quantum efficiency of homodyne detector in the extreme asymmetric
case, and the transmission distance is set to 20 km.
Under this transmission distance, the secret key rate of the CV-MDI-QKD protocols with photon subtraction are lower than that of the original CV-MDI-QKD protocol when the quantum efficiency of homodyne detector values in 0.901 to 1. The reason is that the successful probability of photon subtraction operation in low-channel-loss case is quite low. However, when the detector's quantum efficiency is lower than 0.901, the secret key rate of the protocol with 1-photon subtraction operation become higher than that of the original one. Moreover, when the detector's quantum efficiency is lower than 0.891, the original CV-MDI-QKD protocol will have no secret key. But for the CV-MDI-QKD protocol with 1-photon subtraction, the lower bound of the detector's quantum efficiency for generating secret key in 20 km transmission distance can reach 0.826.
To summarise, the photon subtraction operation can significant reduce the requirement for the quantum efficiency of homodyne detector, and 1-photon subtraction operation performs the best. As the photon subtraction operation can be practically implemented with existing technologies, the CV-MDI-QKD protocol can overcome the limitations of the quantum efficiency of homodyne detector to some extent by adding photon subtraction operation, which gives a feasible scheme for the experimental realization of the CV-MDI-QKD protocol.
\section{Conclusion}\label{Con} 

In this paper, we proposed a method to improve the performance of the CV-MDI-QKD protocol by applying suitable photon-subtraction operations in Alice's entangle source. The proposed method can be implemented based on existing technologies. We give the secret key rates both in the symmetric (\(L_{AC}=L_{BC}\)) and extreme asymmetric (\(L_{BC}=0\)) case. The results show that the photon subtraction operation leads to much longer transmission distance, which precisely make up for the shortcoming of the transmission distance of the CV-MDI-QKD protocol. Moreover, Among all the photon subtraction operations, 1-photon subtraction operation leads to the best performance. Furthermore, the experimental implementation scheme of the CV-MDI-QKD protocol can draw on the ideas of this protocol as it significant lowers the requirement of the quantum efficiency of homodyne detector.
In further research , it would be interesting to combine the photon subtraction with other operations that can enhance the performance of the CV-MDI-QKD protocol, such as noiseless amplification, non-Gaussian state-discrimination detection, etc.

\begin{acknowledgments}
This work was supported the National Basic Research Program of China (Grant No. 2013CB338002), the National Natural Science Foundation of China (Grants No. 11304397, 61332019, 61505261, 61675235, 61605248, 61671287) and the National key research and development program (Grants No. 2016YFA0302600).
\end{acknowledgments}

\begin{appendix}
\section{Detailed calculation of the secret key rate}
After the quantum channel and Charlie's detection, the covariance matrix of \(\rho _{{A_1}{B'_1}}^{PS}\) has the following form

\begin{equation}
\begin{footnotesize}
\begin{array}{l}
\gamma _{{A_1}{{B'}_1}}^{PS} = \left( {\begin{array}{*{20}{c}}
{a{{\rm{I}}_2}}&{c{\sigma _z}}\\
{}&{}\\
{c{\sigma _z}}&{b{{\rm{I}}_2}}
\end{array}} \right)= \left( {\begin{array}{*{20}{c}}
{X{{\rm{I}}_2}}&{\sqrt {\rm{T}} Z{\sigma _z}}\\
{}&{}\\
{\sqrt {\rm{T}} Z{\sigma _z}}&{{\rm{T}}\left( {Y + {\chi _t}} \right){{\rm{I}}_2}}
\end{array}} \right),
\end{array}
\end{footnotesize}
\end{equation}
where \(X\),\(Y\) and \(Z\) are given in Eq. (8), \({\rm I}_2\) is \(2 \times 2\) identity matrix, \({\sigma _z}=diag(1,-1)\). The secure key rate of the CV-MDI-QKD protocol with photon subtraction under one-mode collective Gaussian attack can be calculated with the form
\begin{equation}
{K_{PS}} = {P^{{\rm M^k_1}}}(\beta {I^{PS}_{AB}} - {\chi^{PS} _{BE}}).
\end{equation}
The Shannon mutual information between Alice and Bob, \({I^{PS}_{AB}}\), can be written as [62]
\begin{equation}
{I^{PS}_{AB}} = 2 \times \frac{1}{2}{\log _2}\frac{{{V_{{A_M}}}}}{{{V_{{A_M}|{B_M}}}}},
\end{equation}
where \(V_{{A_M}}=(a+1)/2\) , \({V_{{B_M}}}= (b+1)/2\), and
\begin{equation}
{V_{{A_M}|{B_M}}} = {V_{{A_M}}} - \frac{{{c^2}}}{{4{V_{{B_M}}}}}.
\end{equation}
Then the Shannon mutual information can be obtained as
\begin{equation}
I_{AB}^{PS} = {\log _2}\left( {\frac{{a + 1}}{{a + 1 - {c^2}/(b+1)}}} \right)
\end{equation}

In order to give the lower bound of secure key rate, we adopt that Eve is aware of the mode \(F\) of the untrusted third part Fred, then the state of Eve can be expressed as \(\rho_{EF}\). Since Eve can purify \(\rho _{{A_1}{B'_1}EF}\), \(\rho_{EF}\) can be calculated by \(\rho_{{A_1}{B'_1}}^{PS}\). Then, \({\chi^{PS} _{BE}}\) can be obtained as [60]
\begin{equation}
\begin{array}{lll}
{\chi^{PS} _{BE}} &= S\left( {{\rho _{EF}}} \right) - \int {d{m_B}p\left( {{m_B}} \right)} S\left( {\rho _{EF}^{{m_B}}} \right) \\ \\
&=S\left( {{\rho_{{A_1}{B'_1}}^{PS}}} \right) - S\left( {\rho_{A_1}^{m_{B'_1}}} \right),
\end{array}
\end{equation}
where \(S\) is the Von Neumann entropy of the quantum state \(\rho\), \(m_B\) represents the measurement of Bob. \(\rho _{EF}^{{m_B}}\)  is Eve's state conditional on Bob's measurement result, \(p(m_B)\) is the probability density of the measurement, the covariance matrix of \(\rho _{{A_1}{{B'}_1}}^{PS}\) and \({\rho _{A_1}^{{m_{{B'_1}}}}}\) are  \(\gamma _{{A_1}{{B'}_1}}^{PS}\) and \(\gamma _{{A_1}}^{{m_{{B'_1}}}}\), respectively.
\(S\left( {{\rho_{{A_1}{B'_1}}^{PS}}} \right)\) is a function of the symplectic eigenvalues \(\lambda_{1,2}\) of \(\gamma_{{A_1}{{B'}_1}}^{PS}\) which reads
\begin{equation}
S\left( {{\rho_{{A_1}{B'_1}}^{PS}}} \right)=G[(\lambda_1-1)/2]+G[(\lambda_2-1)/2],
\end{equation}
where
\begin{equation}
G\left( x \right) = \left( {x + 1} \right){\log _2}(x + 1) - x{\log _2}x,
\end{equation}
is the Von Neumann entropy of a thermal state and
\begin{equation}
\lambda _{1,2}^2 = \frac{1}{2}\left( {A \pm \sqrt {{A^2} - 4B^2} } \right),
\end{equation}
where the notations is given as
\begin{equation}
\begin{array}{lll}
A=a^2+b^2-2c^2={X^2} + {{\rm{T}}^2}{\left( {Y + {\chi _{t}}} \right)^2} - 2{\rm{T}}{Z^2},\\ \\
B=ab-c^2= {\rm{T}}\left( {XY + X{\chi _{t}}} - {Z^2}\right).
\end{array}
\end{equation}
\(S\left( {\rho_{A_1}^{m_{B'_1}}} \right)=G[(\lambda_3-1)/2]\) is a function of the symplectic eigenvalues \(\lambda_3\) of \(\gamma _{{A_1}}^{{m_{{B'_1}}}}\), which is given by [62]
\begin{equation}
\gamma _{{A_1}}^{{m_{{B'_1}}}} = {\gamma _{{A_1}}} - \sigma _{{A_1}{{B'}_1}}^TH\sigma _{{A_1}{{B'}_1}},
\end{equation}
where \(H={\left( {{\gamma _{B'_1}}+{{\rm I}_2}} \right)^{-1}}\). Then we can calculate that
\begin{equation}
\begin{array}{lll}
\gamma _{{A_1}}^{{m_{{B'_1}}}} &=a{\rm{I}_2}-c{\sigma _z}(b{\rm{I}_2}+{\rm{I}_2})^{-1}c{\sigma _z}
\\ \\
&=[a-c^2/(b+1)]{\rm{I}_2}
\end{array}
\end{equation}
Thus, we can obtain the symplectic eigenvalues as
\begin{equation}
\lambda_3=a-c^2/(b+1)=X-{\rm{T}} Z^2/[{\rm{T}}(Y+\chi_t)+1].
\end{equation}

\end{appendix}

\end{document}